\documentclass[12pt]{iopart}

\usepackage{graphicx}

\begin{document}


\title{Quadrupolar interactions in Pr compounds:
PrFe$_4$P$_{12}$ and PrBa$_2$Cu$_3$O$_6$}

\author{Annam\'aria Kiss and P. Fazekas}

\address{Research Institute for Solid State Physics and Optics, Budapest 114, P.O.B. 49,
H-1525 Hungary}

\date{\today}

\begin{abstract}
We examine Pr$^{3+}$ crystal field models with near-degeneracy of
the two lowest crystal field levels, where interaction-induced
quadrupolar and dipolar moments are as important as the permanent
moments of the crystal field ground state. We find that the
$\Gamma_1$--$\Gamma_4$ level scheme yields a successful
description of the antiferroquadrupolar ordering of
PrFe$_4$P$_{12}$. For PrBa$_2$Cu$_3$O$_6$, we argue that
quadrupolar interaction is important for understanding the Pr
ordering transition at 11K.
\end{abstract}


\section{Introduction}

The highly degenerate $f$-shells of rare earth and actinide ions
support a great variety of local degrees of freedom: magnetic
(dipolar), quadrupolar, and octupolar. For localized $f$-electron
systems, these (and some of their combinations) are potential
order parameters, leading to rich phase diagrams, as for CeB$_6$
\cite{ceb6}. Field-induced octupoles are known to influence the
shape of the phase boundaries for CeB$_6$ or PrPb$_3$
\cite{prpb3}, but it is an outstanding question whether breaking
time-reversal invariance with purely octupole order is possible;
recent observations on NpO$_2$ indicate that it may be so
\cite{npo2}. The $f$-shell multipolar degrees of freedom may also
interact and mix with those of other electrons: the unusual heavy
fermion behavior of PrFe$_4$P$_{12}$ \cite{Aoki02}, and the exotic
superconducting phases of PrOs$_4$Sb$_{12}$ \cite{izawa} have been
tentatively ascribed to Pr$^{3+}$ quad\-rupole fluctuations.

Detailed understanding has been achieved for CeB$_6$ and some other 
Ce-based systems. The reason is that Ce$^{3+}$ has only a single 
$f$-electron, and therefore microscopic analysis is here the easiest. 
Of particular interest is cubic CeB$_6$ in which  the fourfold
degenerate $\Gamma_8$ ground state supports the full array of 3 dipoles, 5
quadrupoles, and 7 octupoles \cite{ceb6}. A microscopic analysis
\cite{cece} indicates that octupole--octupole interactions should
not be weaker than quadrupole--quadrupole interactions, which
gives rise to a close competition between different kinds of
order. In fact, if all interactions were exactly equal, the model
would be SU(4)-symmetrical, which sparked early theoretical
interest in the subject \cite{ohkawa}. The realistic multipolar
model is not fully symmetrical, but the nearness of the SU(4)
point is still felt in fluctuations \cite{shiina}. Whether the
ground state of an SU(4) model is long range ordered, or it is
rather liquid-like (the SU(4) version of RVB), depends on details
like the lattice structure, and the range of the interaction
\cite{penc}. The approximate realization of high symmetries like
SU(4) is of current interest for $d$- and $f$-electron systems
alike.

We expect that the result that various multipolar couplings are of
the same order of magnitude, holds generally for light rare
earths. The question arises what can we say about $4f^2$ systems.

Pr compounds show varied behavior. A few examples: PrPb$_3$
undergoes an antiferroquadrupolar transition but it does not order
magnetically, though both dipolar and octupolar interactions are
thought to be important \cite{prpb3}. PrB$_6$ has two magnetic
phase transitions but the magnetic anisotropy is explained by
quadrupolar interactions \cite{prb6}. PrBa$_2$Cu$_3$O$_{7-x}$ is
the only member of the group (RE)Ba$_2$Cu$_3$O$_7$
 (RE=rare earth), which is not a high-$T_C$
superconductor \cite{rado}. It is suspected that this
is related to the little-understood Pr ordering transition at
17K, but the connection is not clear. The parent Mott insulator
PrBa$_2$Cu$_3$O$_6$ has a similar phase transition at 11K
\cite{uma}.

\section{Pr systems}

In most compounds, Pr is trivalent. The existence of the dioxide
PrO$_2$ shows that it can also be tetravalent \cite{pro2}, though
this is rare. The question of the valence state becomes crucial
whenever the nature, or the driving force, of Pr ordering is in
doubt. For Pr-filled skutterudites: the formation of a heavy
Fermi sea, as seen in PrFe$_4$P$_{12}$ and in PrOs$_4$Sb$_{12}$,
is most likely to involve the hybridization of Pr $f$-states with
conduction band states, and consequently deviation from strictly
integral valence.

These basic doubts notwithstanding, we propose to study the
interplay of dipolar and quadrupolar interactions in localized
$f$-electron models, and apply the results to PrFe$_4$P$_{12}$ and
PrBa$_2$Cu$_3$O$_6$. We assume the integer valence state
Pr$^{3+}$, and a Hund's rule ground state with $J=4$. We note that
a similar approach was successful in the case of CeB$_6$
\cite{ceb6}, though the system is known to possess the features of a
dense Kondo system. We envisage a similar study of Pr systems.

Pr$^{3+}$ is a non-Kramers ion. It follows that a unique ground
state can be reached without breaking time-reversal invariance,
either by having a singlet $\Gamma_1$ crystal field ground state
to begin with, or via the ordering of quadrupolar moments.

\subsection{The $\Gamma_1$--$\Gamma_4$ scenario for PrFe$_4$P$_{12}$}

PrFe$_4$P$_{12}$ undergoes a second order transition at $T_{\rm
tr}=6.51{\rm K}$, which is manifested in a susceptibility spike,
and a $\lambda$-shaped anomaly of the specific heat \cite{Aoki02}.
Though at first thought to be antiferromagnetic, the ordered phase
is now known to be antiferroquadrupolar. Upon switching on a
magnetic field, the transition temperature decreases. The
character of the transition switches to first order at the
tricritical point $T_{\rm tri}\approx 5{\rm K}$, $H_{\rm
tri}\approx 2{\rm T}$, and the ordered phase is completely
suppressed at $\sim 4{\rm T}$ \cite{Aoki02}. These parameters
depend on the direction of the applied field. However, here we
consider only the case of a field applied along the (100)
direction.

The question whether the $f$-states are itinerant or localized can
be posed again. PrFe$_4$P$_{12}$ is one of the few Pr-based heavy
fermion systems \cite{Aoki02}. Though having an even number of
$f$-electrons per site makes the distinction between large and
small Fermi surfaces difficult, we take the view that the large
Fermi sea of heavy $f$-electrons competes with a state in which
the conduction electrons build a small Fermi sea, and multipolar
inter-shell interactions are switched on \cite{shiba,mh,coleman}.
This is supported by inelastic neutron scattering finding that the
crystal field excitations become sharply defined when the
temperature is lowered below $T_{\rm tr}$. It follows that we may use a
localized $f$-electron model to describe the ordered phase and the
vicinity of the phase boundary. We use the crystal field states of
the octahedral group $O_h$. We note that the icosahedron of the
twelve P atoms surrounding a Pr site gives a novel tetrahedral
component to the crystal field potential \cite{take}, but we
assume that using cubic symmetry labelling  is an acceptable
approximation.

Since the local order parameter is one of the quadrupolar
moments, it may look evident that the crystal field ground
state should carry a (spontaneous) quadrupolar moment. However, an
analysis of the high-$T$ behavior of the magnetization curves shows 
that the level scheme (ground state)--(first excited
state) may be either of the following three:
$\Gamma_1$--$\Gamma_4$, $\Gamma_1$--$\Gamma_5$, or
$\Gamma_3$--$\Gamma_4$ \cite{Aoki02}. Only the last, with the
$\Gamma_3$ ground state, carries an unfrozen quadrupolar moment
which can be ordered by switching on (arbitrarily small) intersite
interactions. It was also shown that this choice is consistent
with a symmetry analysis of the structural distortion accompanying
the antiferro-quadrupolar ordering \cite{curnoe}. This latter
argument relies only on the assumption of the $\Gamma_3$ ground
state, and does not refer to the nature of the first excited
state.

Here we show that the alternative $\Gamma_1$--$\Gamma_4$ scheme is
also capable to account for most of the observed static properties of
PrFe$_4$P$_{12}$. We list the crystal field states
\begin{eqnarray}
\Gamma_1 & = & \sqrt{5/24}(|4\rangle + |-4\rangle) +
\sqrt{7/12}|0\rangle\nonumber \\
\Gamma_4^{\pm} & = & (1/4) (|3\rangle \pm |-3\rangle +
\sqrt{7}[\pm|1\rangle + |-1\rangle])\nonumber \\
 \Gamma_4^{0}  & = & \sqrt{1/2}  (|-4\rangle - |4\rangle).\label{eq:gamma4}
\end{eqnarray}
where we use the quadrupolar eigenstates $\Gamma_4^{\pm}$ in the
representation $\Gamma_4$. Since the ground state carries no kind
of moment, the ordered quadrupolar moment must be
interaction-induced, implying that the interaction has to exceed a
threshold value of the order of the $\Gamma_1$--$\Gamma_4$
splitting. Taking it for granted that intersite interactions mix
the $\Gamma_4$ states with $\Gamma_1$, the possible local order
parameters are those appearing in the decomposition
\begin{equation}
(\Gamma_1\oplus\Gamma_4)\otimes(\Gamma_1\oplus\Gamma_4) =
2\Gamma_1+\Gamma_3+3\Gamma_4+\Gamma_5\, .
\label{eq:decomp1}
\end{equation}
The system is capable of either  $\Gamma_3$ or $\Gamma_5$ type
quadrupolar ordering, or any (dipolar or octupolar) kind of
$\Gamma_4$ order. In principle, we have to allow for all the
corresponding intersite interactions. However, we neglect the
octupoles. We assume $\Gamma_3$-type quadru\-polar order and,
recalling that the pair interaction has only tetragonal symmetry,
and so ${\cal O}_2^2$ and ${\cal O}_2^0$ are not equivalent, we
arbitrarily keep only the ${\cal O}_2^2=J_x^2-J_y^2$ term. The 
following  mean-field-decoupled hamiltonian includes also a
dipole-dipole coupling:
\begin{eqnarray}
\mathcal H & = & \mathcal H_{CF}+\mathcal H_{\rm Zeeman}+\mathcal
H_{\rm quad}+\mathcal H_{\rm dipole}\nonumber
\\
&=& \Delta
\sum_{n=0,\pm}\left|\Gamma_{4}^n\right\rangle\left\langle\Gamma_{4}^n\right|
-g\mu_{\rm B}\mathbf H\mathbf J-zQ{\left\langle {\mathcal
O}_{2}^{2}\right\rangle}_{B(A)} {\mathcal O}_{2}^{2}
-zI{\left\langle\mathbf J\right\rangle}_{B(A)}{\cdot}\mathbf J
\label{eq:meanf}
\end{eqnarray}
where we allowed for two-sublattice (A and B) order on the bcc
lattice ($z=8$). $g=4/5$ is the Land\'e factor, $Q$ the
quadrupolar, and $I$ the dipolar coupling constant. We mention that
a simplified interaction, with quadrupolar coupling only, was
treated in a previous communication \cite{amk}. However, the
inclusion of ${\cal H}_{\rm dipole}$ is important for getting a
better fit to the experimental results, in particular the
susceptibility.

\begin{figure}[ht]
\begin{center}
\includegraphics[width=5cm]{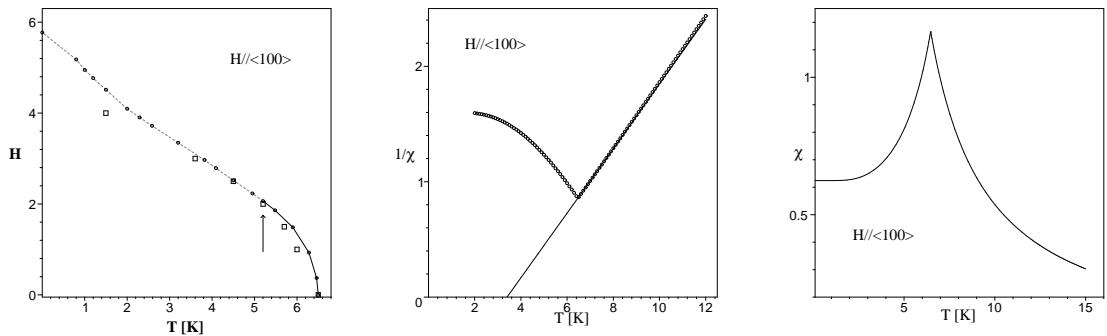}
\includegraphics[width=5cm]{fazekas_fig_1_middle.ps}
\includegraphics[width=5cm]{fazekas_fig_1_right.ps}
\end{center}
\caption{{\sl Left}: The phase boundary in the $H$--$T$ plane
($H\parallel{\hat x}$). Continuous line: second order, dashed
line: first order transitions. Open circles: calculated, open
rectangles: measured (after \protect{\cite{Aoki02}}). {\sl Middle}
and {\sl right}: the inverse susceptibility $1/\chi$, and the
susceptibility $\chi$, as a function of $T$. $\chi$ is in units of
$\mu_{\rm B}/{\rm Tesla}$/(Pr site).}\label{fig:skut_sus}
\end{figure}

The diagonalization of (\ref{eq:meanf}) in the basis
(\ref{eq:gamma4}) is straightforward. We show only the results. We
have found that the (ferromagnetic) dipolar coupling $I=106{\rm
mK}$, the antiferro-quadrupolar coupling $Q=-9.5{\rm mK}$, and the
crystal field splitting $\Delta=3{\rm K}$ give good overall
agreement with the observations (in case $Q$ appears implausibly
small, check that ${\cal O}_2^2$ has big matrix elements). With
these parameters, the only phase transition is the onset of
antiferro-quadrupolar order at $T_{\rm tr}=6.5{\rm K}$. The
transition is gradually suppressed by an external magnetic field
(Fig.~\ref{fig:skut_sus}, left). We locate the tricritical point
at $T_{\rm tri}\approx 5{\rm K}$, $H_{\rm tri}\approx 2{\rm T}$,
in agreement with experiment. The same set of parameters gives
also the observed Weiss temperature $\Theta=3.6{\rm K}$ from the
intercept in the inverse susceptibility (Fig.~\ref{fig:skut_sus},
middle).

We note that our estimate of the crystal field splitting
$\Delta\approx 3{\rm K}$ is substantially smaller than the one
quoted in \cite{Aoki02}; considering the scale of $T_{\rm tr}$, we
have a quasi-quadruplet. However, this is not necessarily
unreasonable: a calculation indicates a region where the gaps
become very small (H. Harima, private communication).

\begin{figure}[ht]
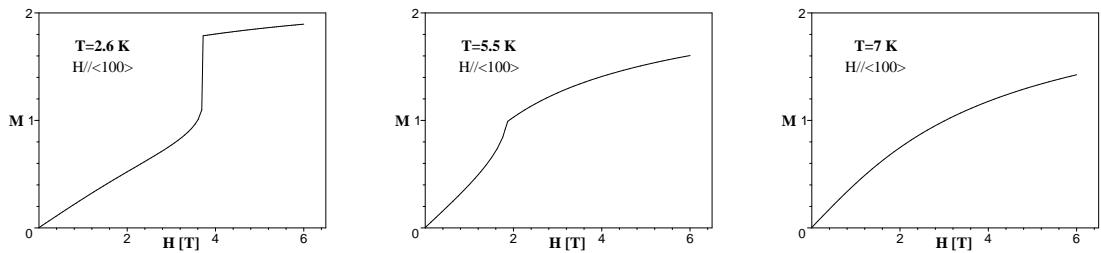

\begin{center}
\includegraphics[width=5cm]{fazekas_fig_2_left.ps}
\includegraphics[width=5cm]{fazekas_fig_2_middle.ps}
\includegraphics[width=5cm]{fazekas_fig_2_right.ps}
\end{center}
\caption{Magnetization curves for the $\Gamma_1$--$\Gamma_4$
model, below (left) and above (middle) of the tricritical
temperature, and in the disordered phase (right). Parameter values
are the same as in
Fig.~\protect\ref{fig:skut_sus}.}\label{fig:skutmag}
\end{figure}

To conclude our review of the results from the
$\Gamma_1$--$\Gamma_4$ scheme\footnote{We did not make 
a comparable study of the full range of models, so we could not claim that 
the present model gives best agreement with experiments. Rather, we wished 
to demonstrate that the $\Gamma_1$--$\Gamma_4$ scheme is not obviously 
excluded, in spite of the absence of ground state quadrupolar moments.}, 
we show the magnetization curves
for three representative temperatures (Fig.~\ref{fig:skutmag}). In
the region of first order transitions below the tricritical
temperature, one finds a sharp metamagnetic transition which
becomes smoother in the regime of continuous transitions. These
results show a good overall resemblance to the measured curves
\cite{Aoki02}.

\subsection{The role of quadrupolar coupling in PrBa$_2$Cu$_3$O$_6$}

All the (RE)Ba$_2$Cu$_3$O$_7$ systems, with the exception of
RE=Pr, are $T_C\approx 90{\rm K}$ superconductors. Whenever the
RE$^{3+}$ ion has a magnetic moment, they also undergo RE
antiferromagnetic ordering at $T_{\rm N}=1-2{\rm K}$. Analogous
results hold for the parent Mott insulators (RE)Ba$_2$Cu$_3$O$_6$,
where most $T_{\rm N}$ follow de Gennes scaling \cite{degennes},
the peak belonging to RE=Gd with $T_{\rm N}\approx 2.2{\rm K}$
\cite{feri}. This is well understood on the basis of the $I_{\rm
Gd-Gd}=156{\rm mK}$ meaured by ESR \cite{feri}. We would expect
that for RE=Pr, 
\[\left.T_{\rm N}({\rm Pr})\right|_{\rm de
Gennes}=\frac{\left\{ (g-1)^2J(J+1)\right\}_{\rm Pr}}{\left\{
(g-1)^2J(J+1)\right\}_{\rm Gd}}{\cdot}T_{\rm tr}({\rm Gd})\approx
0.1{\rm K}\; .
\] 
Instead,
magnetic ordering and, at the same time, tetragonal-to-orthorombic
distortion, is observed at $T_{\rm N}\approx 17{\rm K}$ for
PrBa$_2$Cu$_3$O$_7$, and $T_{\rm N}\approx 11{\rm K}$ for
PrBa$_2$Cu$_3$O$_6$ \cite{uma}, exceeding the de Gennes estimate by 
two orders of magnitude. There must be a mechanism for
producing substantially stronger exchange, or a different kind of
ordering, or both.

ESR on Gd:PrBa$_2$Cu$_3$O$_6$ gives the intersite exchange $I_{\rm
Pr-Gd}=-140{\rm mK}$ \cite{unpub}. De Gennes scaling would lead us
to expect this sign, but also an absolute value of about a factor
of 4 smaller. The greater spatial extent of Pr $4f$ orbitals is
likely to explain the enhancement of $I$; correspondingly, we
would expect that $I_{\rm Pr-Pr}$ is even more enhanced, but would
probably still fall short of accounting for the observed $T_{\rm
N}$. Here we suggest that the presence of quadrupolar interactions
may explain several features of the magnetic behavior of
PrBa$_2$Cu$_3$O$_6$, and also an additional enhancement of $T_{\rm
N}$.

All previous works agree that in the level scheme of Pr$^{3+}$
ions, a low-lying quasi-triplet consisting of the tetragonal
doublet
\begin{equation}
\Gamma_t^{\pm} = \alpha|\pm3\rangle - \beta|\mp 1\rangle
\label{eq:doub}
\end{equation}
and the singlet
\begin{equation}
\Gamma_t^{0} = \sqrt{1/2}\;(|2\rangle - |-2\rangle)
\label{eq:sing}
\end{equation}
is well separated from the remaining six $J=4$ states, and
therefore suffices for modelling (as far as the Pr sites are
concerned) all low-$T$ phenomena \cite{hilscher,uma}. With
$\alpha=\sqrt{7/8}\approx 0.9354$, (\ref{eq:doub}) and
(\ref{eq:sing}) would constitute the cubic $\Gamma_5$ triplet, but
our fits yield $\alpha\approx 0.943$, indicating a slight
admixture from the doublet derived from $\Gamma_4$. Let us observe
that the doublet carries both $J^z$ dipole, and ${\cal O}_2^2$ (or
alternatively ${\cal O}_{xy}$) type quadrupolar moment. If the
doublet is the crystal field ground state, the degeneracy can be
resolved either by magnetic, or by quadrupolar, ordering.

 Using (\ref{eq:doub})--(\ref{eq:sing}), and following the procedure of \cite{morin},
 we fitted the high-$T$ magnetization curves, and found
 that the dipole--dipole coupling has a strong planar anisotropy,
 and there is also a substantial ferroquadrupolar coupling of the
 ${\cal O}_2^2$ moments \cite{unpub}. The latter finding may lead
 us to ask whether the observed transition is perhaps purely of
 quadrupolar nature; however, neutron scattering shows that $T_{\rm
N}$ is indeed a N\'eel temperature, with the $T<T_{\rm N}$ ordered
moments strongly tilted out of the tetragonal $c$-direction
 \cite{boothroyd}. Within the crystal field model, it is rather
 mysterious why the system does not take advantage of the
 permanent $c$-axis moments of the doublet (\ref{eq:doub}), and
 chooses instead $ab$-plane moments which have to be
 interaction-induced.

The mean field hamiltonian acting on the quasi-triplet is similar
to (\ref{eq:meanf})
\begin{eqnarray}
\mathcal H & = & \mathcal H_{CF}+\mathcal H_{\rm Zeeman}+\mathcal
H_{\rm quad}+\mathcal H_{\rm dipole}\nonumber
\\
&=& \Delta
\left|\Gamma_{t}^0\right\rangle\left\langle\Gamma_{t}^0\right|
-g\mu_{\rm B}\mathbf H\mathbf J-zQ{\left\langle {\mathcal
O}_{2}^{2}\right\rangle} {\mathcal O}_{2}^{2} -zI{\left\langle
J_x\right\rangle}_{B(A)}J_x \label{eq:meanfb}
\end{eqnarray}
where we allow for two-sublattice antiferromagnetism, but assume
ferroquadrupolar coupling. In view of the bilayer structure of
PrBa$_2$Cu$_3$O$_6$, we take $z=4$. From our susceptibility fits,
we estimate $\Delta\approx 20{\rm K}$.

We note that a closely related three-state problem was considered
by L\'{\i}bero and Cox \cite{libero}. They assume hexagonal
symmetry, so their choice of the singlet and the doublet is quite
different from ours, but nevertheless, when worked out fully, the
results of their mean field theory should basically correspond to
what we find. However, we calculate different quantities (such as
the susceptibilities), and emphasize different aspects.

\begin{figure}[ht]
\begin{center}
\includegraphics[width=5.0cm]{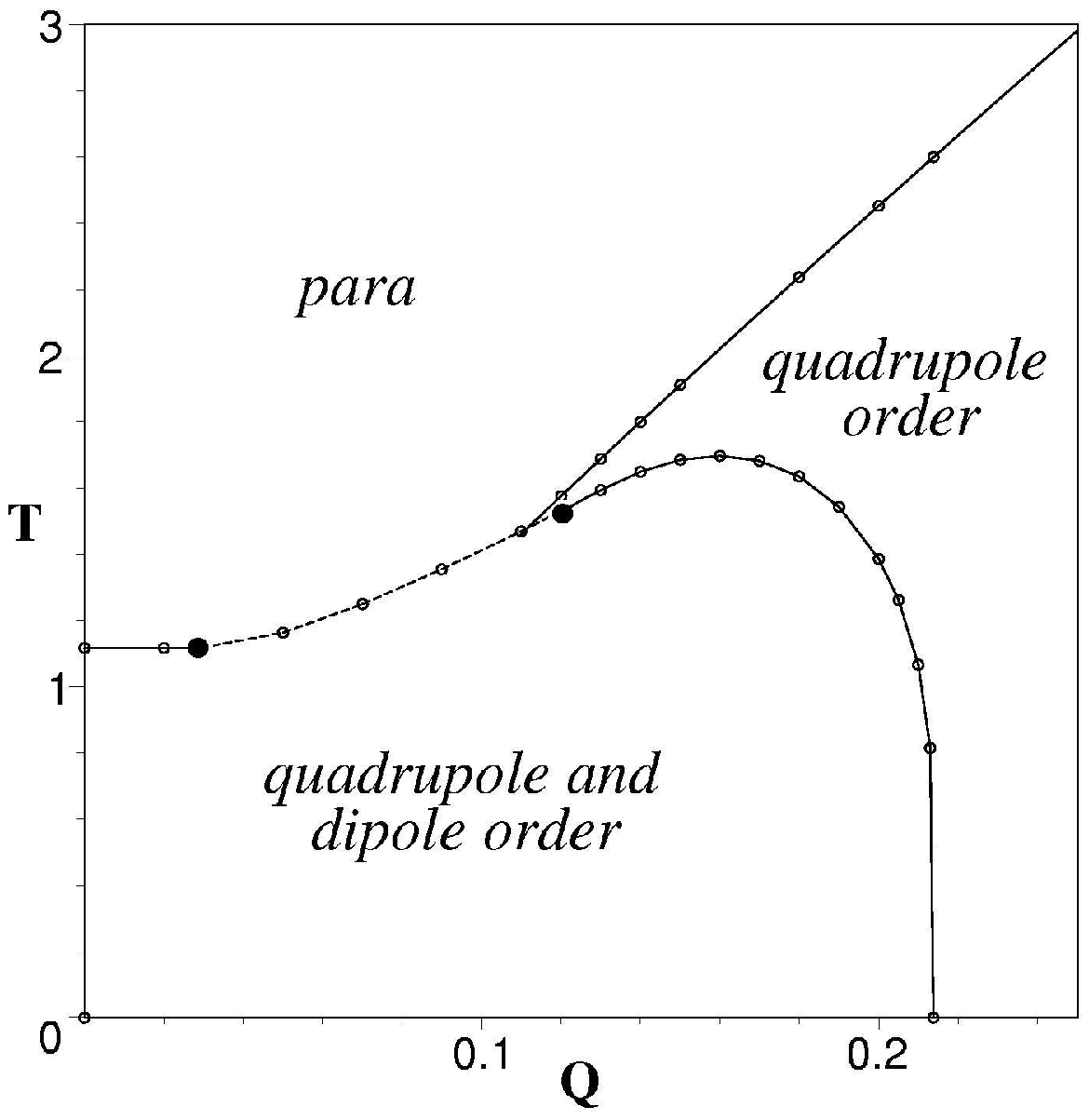}
\includegraphics[width=5.0cm]{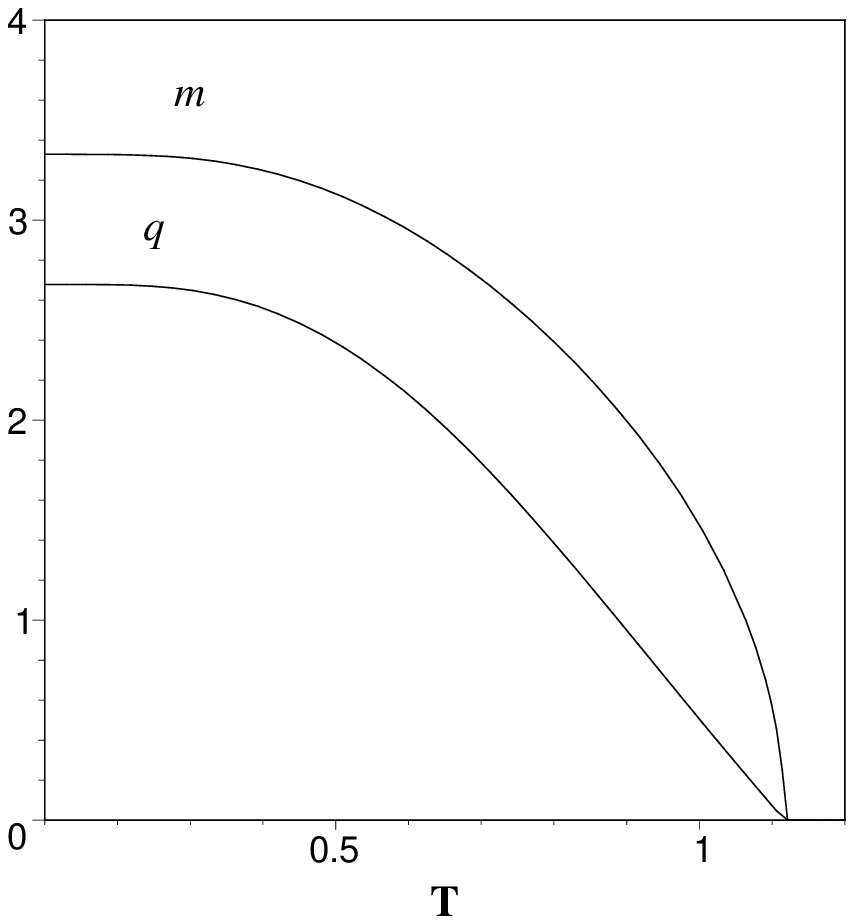}
\includegraphics[width=5.0cm]{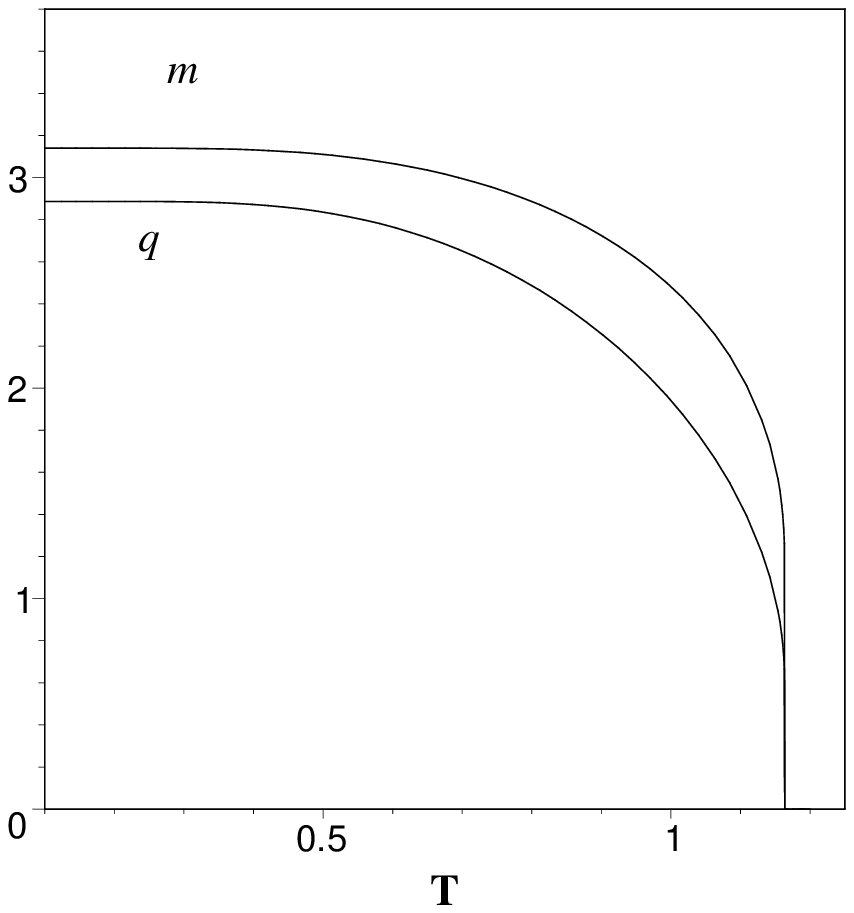}
\end{center}
\caption{{\sl Left}: Transition temperatures as a function of the
quadrupolar coupling $Q$[K], for $I=0.2$K and $\Delta=2{\rm K}$.
Circles: calculated points; the dashed (first order)
and continuous (second order) lines are drawn for
convenience. Filled circles: tricritical points. 
{\sl Middle and right}: magnetic (m) and quadrupolar
(q) order parameters for $Q=0$ and $Q=0.05$, resp.}
\label{fig:prob6a}
\end{figure}

Fig.~\ref{fig:prob6a} (left) shows a representative phase diagram.
At $Q=0$, the system gains energy from antiferromagnetic ordering,
with the moments lying in the ${\pm}x$-direction. However, it is
easy to convince ourselves that at the same time, the system
develops quadrupolar order: $J^x$ connects $\Gamma_t^0$ to
$\Gamma_t^+ -\Gamma_t^-$ (which is one of the quadrupolar
eigenstates), but not to $\Gamma_t^+ +\Gamma_t^-$ (which would be
the other). Thus, for zero or weak $Q$, the primary order
parameter is $\langle J_x\rangle$, while $\langle{\cal
O}_2^2\rangle$ is a secondary order parameter
(Fig.~\ref{fig:prob6a}, middle). At sufficiently strong $Q$, both
interactions are important, and the two orders appear at a first
order transition (Fig.~\ref{fig:prob6a}, right). The regime of 
first-order transitions is bounded by a lower, and an upper, tricritical 
point (big black dots in Fig.~\ref{fig:prob6a}, left). Above a 
$\Delta$-dependent threshold value of $Q/I$, pure quadrupolar order
sets in at a higher critical temperature.  
From this point on, the quadrupolar splitting is added to
the crystal field splitting, eventually suppressing 
magnetism. Note, however, that for a wide range of $Q/I$, {\sl
magnetism is assisted by quadrupolar interactions}, in the sense
that the magnetic transition temperature increases with $Q/I$.
This holds also for part of the regime where the two transitions
are distinct.

\begin{figure}[ht]
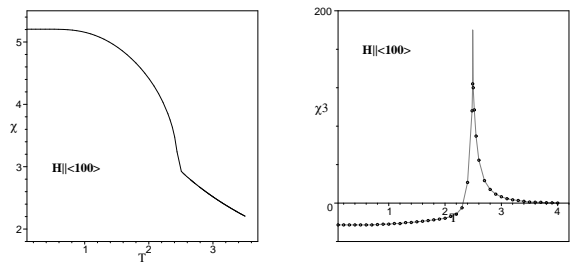

\begin{center}
\includegraphics[width=4cm]{fazekas_fig_4_left.ps}
\includegraphics[width=4cm]{fazekas_fig_4_right.ps}
\end{center}
\caption{The $T$-dependence of linear ({\sl left}) and nonlinear
susceptibility ({\sl right})  for purely ferroquadrupolar
ordering. $\chi_3$ diverges at the onset of quadrupolar
order.}\label{fig:qsusc}
\end{figure}

\begin{figure}[ht]
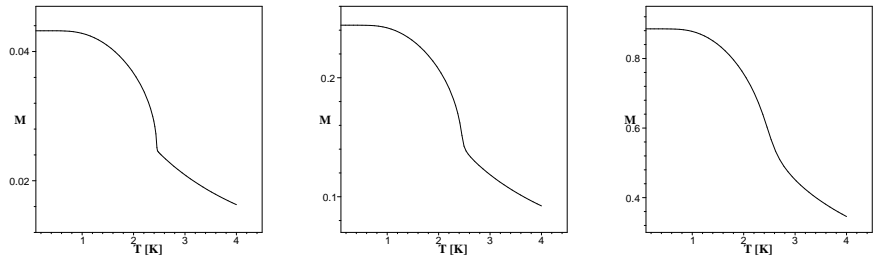

\centering
\includegraphics[totalheight=3.8cm]{fazekas_fig_5_left.ps}
\includegraphics[totalheight=3.8cm]{fazekas_fig_5_middle.ps}
\includegraphics[totalheight=3.8cm]{fazekas_fig_5_right.ps}
\caption{Magnetization ($\mu_{\rm B}$/Pr) versus temperature for
the $H\parallel (1,0,0)$ magnetic fields $0.01\rm T$ ({\sl left}),
$0.1\rm T$ ({\sl center}) and $0.4\rm T$ ({\sl right}).
}\label{fig:qmagn}
\end{figure}

A remarkable aspect of Pr ordering is its sensitivity to magnetic
field, particularly for $H\perp (001)$ \cite{naroz,unpub}.
Quadrupolar interactions are known to have a strong signature in
non-linear magnetic response \cite{morin}. We show in
Fig.~\ref{fig:qsusc} the results for $I=0$, $Q=0.2$, $\Delta=2$,
for the field in the $ab$ plane. The upward curvature of $\chi$,
and the divergence of $\chi_3$ at the transition are reminiscent
of the observed behavior. Fig.~\ref{fig:qmagn} shows the
$T$-dependence of the magnetization for several fields $H\parallel
(100)$. It is apparent that the sharp $H=0$ anomaly is quickly
smeared out in higher fields, similar to the observed behavior of
PrBa$_2$Cu$_3$O$_6$.

To conclude, we have examined models where the local Hilbert space
is spanned by the states from the lowest two crystal field levels.
The near-degeneracy of two levels has interesting consequences
when the splitting is comparable to both intersite intractions,
and laboratory magnetic fields. We found that the induced
quadrupolar moment scenario gives a good understanding of the
static properties of PrFe$_4$P$_{12}$, and that the inclusion of
quadrupolar interaction is helpful in understanding the non-linear
magnetic behavior of PrBa$_2$Cu$_3$O$_6$.


We have been supported by the Hungarian grants OTKA T
038162 and TS 040878. Support from a JSPS--HAS exchange program enabled 
P.F. to participate at ASR2002.


\end{document}